\documentstyle[sprocl]{article}
\begin{document}
\renewcommand\floatpagefraction{.9}
\renewcommand\topfraction{1}
\renewcommand\bottomfraction{1}
\renewcommand\floatsep{12pt}
\renewcommand\textfraction{0}
\renewcommand\intextsep{12pt}
\def\gtorder{\mathrel{\raise.3ex\hbox{$>$}\mkern-14mu
             \lower0.6ex\hbox{$\sim$}}}
\def\ltorder{\mathrel{\raise.3ex\hbox{$<$}\mkern-14mu
             \lower0.6ex\hbox{$\sim$}}}
\input psfig

\title{SOLAR NEUTRINOS: WHERE WE ARE}
\author{JOHN BAHCALL}
\address{Institute for Advanced Study, Princeton, NJ 08540}
\maketitle
\abstracts{This talk compares standard model 
predictions for solar neutrino experiments with the results of 
actual observations. Here `standard model' means the combined standard
model of minimal electroweak theory plus a standard solar model.
I emphasize the importance of recent analyses
in which the neutrino fluxes are treated as free parameters,
independent of any constraints from solar models, and the stunning 
agreement between the predictions of standard solar models and 
helioseismological measurements.}

\section{Introduction}
\label{intro}
Joe Taylor mentioned in his beautiful 
preceding review of pulsar phenomena that
the discussion of pulsars has a long history at the Texas conferences.
I want to add a footnote to his historical remarks: the subject of
solar neutrinos has an even longer history in this context.  Both Ray
Davis and I gave invited talks~\cite{Bahcall64,Davis64a}
 on solar neutrinos at the 2nd Texas
Conference, which was held in Austin, Texas in December 1964.

Solar neutrino research has now achieved the primary goal that was 
discussed in the 1964 Texas Conference, namely, the 
detection of solar neutrinos.  This detection establishes empirically
that the sun shines by fusing light nuclei in its interior.

The subject of solar neutrinos 
is  entering a new phase in which
large electronic detectors will yield vast amounts of diagnostic data.
These new experiments~\cite{Arp92,Takita93,McD94}, which  will be
described by Professor Totsuka in the following talk,
will test the prediction of the minimal standard
electroweak theory~\cite{Glashow61,Wein67,Salam68} 
that essentially nothing happens to electron type
neutrinos after they are created by nuclear fusion reactions in the
interior of the sun.

The four pioneering experiments---chlorine~\cite{Davis64,Davis94} 
Kamiokande~\cite{Suzuki95} 
GALLEX~\cite{Ansel95}  and SAGE~\cite{Abdur94}---have 
all observed neutrino fluxes with intensities that are within a
factors of a few of  
those predicted by standard solar models. 
Three of the experiments (chlorine, GALLEX, and SAGE) are
radiochemical and each radiochemical experiment  measures
 one number, the total rate at which
neutrinos above a fixed energy threshold (which depends upon the
detector)  are captured.  
The sole electronic (non-radiochemical) 
detector among the initial experiments,
Kamiokande, has shown 
that the neutrinos come from the sun,
by measuring the recoil directions of the  electrons scattered by
solar neutrinos.
Kamiokande has also demonstrated 
that the observed neutrino energies 
are consistent with 
 the range of energies expected on the basis of the standard solar model.

Despite continual refinement of solar model calculations of
neutrino fluxes over the past 35 years (see, e.g., the collection of
 articles reprinted in the book edited by 
Bahcall, Davis, Parker, Smirnov, and Ulrich~\cite{BDP95}),
the discrepancies between 
observations and calculations have gotten worse with time.  All four
of the pioneering solar neutrino experiments yield event rates that
are significantly less than predicted by standard solar models.

This talk is organized as follows.
I first discuss in section~\ref{threeproblems} the three solar
neutrino problems. Then I review 
in section~\ref{lasthope}
the recent work by Heeger and
Robinson~\cite{HR96} which 
treats the neutrino fluxes as free parameters and 
shows that the solar neutrino problems 
cannot be resolved within the context of minimal standard electroweak
theory unless     solar neutrino experiments are incorrect.
Next I discuss in section~\ref{helioseismology} the stunning agreement
between the values of the sound velocity calculated from standard
solar models and the values obtained from helioseismological
measurements. Finally, in section~\ref{versus} I compare the success
of the MSW neutrino mixing hypothesis with the success of the 
solar $^3$He mixing
hypothesis recently discussed by Cumming and Haxton\cite{Haxton96}.

See
http://www.sns.ias.edu/\raise2pt\hbox{${\scriptstyle\sim}$}jnb 
for further information about solar neutrinos, including viewgraphs,
preprints, and numerical data.

\section{Three Solar Neutrino Problems}
\label{threeproblems}

I will first compare  the predictions of the combined
standard model with the results of the  operating solar neutrino
experiments.    
By `combined' standard model, I mean the predictions of the standard
solar model and the predictions of the minimal electroweak theory.
We need a solar model to tell us how many neutrinos of what energy 
are produced in the sun and we need electroweak theory to tell us how
the number and flavor content of the neutrinos are changed as they
make their way from the center of the sun to detectors on earth.

We will see that this comparison leads to three
different discrepancies between the calculations and the observations,
which I will refer to as the three solar neutrino problems.

\begin{figure}[htb]
\hglue.3in{\psfig{figure=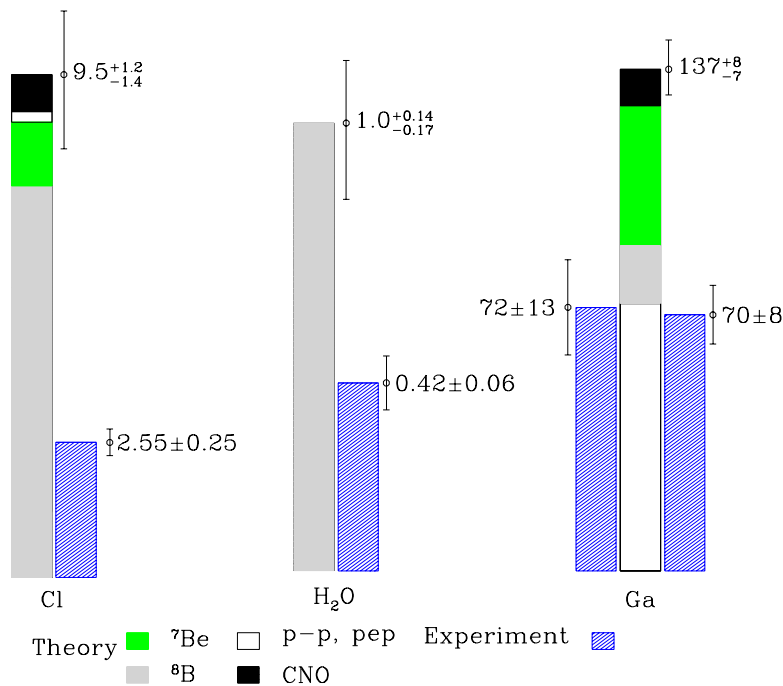,width=4.2in}}
\caption[]{\baselineskip=10pt{\footnotesize Comparison of measured 
rates and 
standard-model predictions
for four solar neutrino experiments\label{compare}.}\smallskip}
\end{figure}

Figure~\ref{compare} shows  the measured and the calculated event
rates in the four ongoing solar neutrino experiments.  This figure
reveals three discrepancies between the 
experimental results and the expectations based
upon the combined standard model.  As we shall see, only
the first of these discrepancies depends sensitively upon
predictions of the standard solar model.

\subsection{Calculated versus Observed Absolute Rate}
\label{firstproblem}

The first solar neutrino experiment to be performed was the chlorine
radiochemical experiment, which detects electron-type neutrinos that
are  more energetic 
than $0.81$ MeV.  After more than
25 years of the operation of this experiment, the measured event
rate is $2.55 \pm 0.25$ SNU, which is a factor $\sim 3.6$ less than
is predicted by the most detailed theoretical calculations,
$9.5_{-1.4}^{+1.2}$ SNU~\cite{BP95,BPBJCD}.  
A SNU is a convenient unit to describe the
measured rates of solar neutrino experiments: $10^{-36}$ interactions
per target atom per second. 
Most of the predicted rate in the chlorine experiment is
from the rare, high-energy $^8$B neutrinos, although the $^7$Be
neutrinos are also expected to contribute significantly.  According to
standard model calculations, the $pep$ neutrinos and the CNO neutrinos 
(for simplicity not discussed here)
are expected to contribute less than 1 SNU to the
total event rate.

This discrepancy between the calculations and the observations for the
chlorine experiment was, for more than two decades, the only solar
neutrino problem. I shall refer to the chlorine disagreement 
as the ``first'' solar neutrino
problem.

\subsection{Incompatibility of Chlorine and Water (Kamiokande) 
Experiments}

The second solar neutrino problem results from a comparison of the 
measured event rates in the chlorine experiment and in the Japanese
pure-water experiment,  Kamiokande.  The water experiment detects
higher-energy neutrinos, those with energies  above $7$ MeV,
by neutrino-electron scattering: $\nu ~+~e ~\longrightarrow
  \nu' ~+~e'.$   According to the standard solar model, 
\hbox{$^{8}$B} beta decay is the only important 
source of these higher-energy neutrinos. 

The Kamiokande experiment shows that the observed neutrinos come from
the sun. 
The  electrons that are scattered by the incoming neutrinos recoil
predominantly  in the direction of the sun-earth vector;  the
relativistic electrons are observed by the 
Cherenkov radiation they produce in the water detector.

In addition, the Kamiokande 
experiment measures
the energies of individual scattered electrons and therefore
provides information about the energy spectrum of the incident 
solar neutrinos. The observed 
spectrum of electron recoil energies 
is consistent with that expected from $^8$B neutrinos.
However,  small angle scattering of the recoil  electrons in the water
prevents the angular distribution from being determined well on an
event-by-event basis, which limits  the constraints the experiment
places on the incoming neutrino energy  
spectrum.

The event rate in the Kamiokande experiment is
determined by the same high-energy $^8$B neutrinos that are expected,
on the basis of the combined standard model,
to dominate the event rate in the chlorine experiment.
I have  shown\cite{Bahcall91} that solar physics changes   
the shape of the \hbox{$^{8}$B} neutrino spectrum by less than  1 part
in $10^5$~. 
Therefore, we can calculate the rate in the chlorine experiment that
is produced by  
the \hbox{$^{8}$B} neutrinos observed
in the Kamiokande experiment (above 7  MeV).
This partial (\hbox{$^{8}$B}) rate in the chlorine experiment 
is $3.2 \pm
0.45$ SNU, which exceeds the total observed chlorine 
rate of $2.55 \pm 0.25$ SNU. 

Comparing the rates of the
Kamiokande and the chlorine experiments, one finds that the net
contribution to the chlorine experiment from the $pep$, $^{7}$Be, and CNO
neutrino sources is negative: $-0.66 \pm 0.52$ SNU.
 The standard model calculated rate from $pep$, $^7$Be, and CNO neutrinos is
1.9~SNU.  The apparent incompatibility of the chlorine
and the Kamiokande
experiments is the ``second'' solar neutrino problem.
The inference that is often made from this comparison is that the
energy spectrum of ${\rm ^8B}$ neutrinos is changed from the standard
shape by physics not included in the simplest version of the standard
electroweak model.

\subsection{Gallium Experiments: No Room for $^{7}$Be Neutrinos}
\label{galliumproblem}

The results of the 
gallium experiments, GALLEX and SAGE,
constitute the third solar neutrino problem.
The average observed  rate in these two experiments is $70.5 \pm 7$ 
SNU, which
is fully accounted for in the standard model by the 
theoretical  rate of $73$ SNU
that is calculated to come from the basic $p$-$p$ and $pep$ neutrinos
(with only a 1\% uncertainty in the standard solar model $p$-$p$ flux).
The \hbox{$^{8}$B} neutrinos, which are observed above $7.5$ MeV 
in the Kamiokande experiment, must also contribute to the gallium
event rate. 
Using the standard shape for the spectrum of ${\rm ^8B}$
neutrinos and normalizing to the rate observed in Kamiokande, 
${\rm ^8B}$ contributes  
another $7$ SNU,
unless something happens to the lower-energy  neutrinos
after they are created in the sun. (The predicted contribution is
16~SNU on the basis of the standard model.)  
Given the measured rates in the
gallium experiments, there is no room for the additional $34 \pm 4$
SNU that
is expected from $^{7}$Be neutrinos on the basis of 
standard solar models.

The seeming exclusion of everything but $p$-$p$ neutrinos in the gallium
experiments is the ``third'' solar neutrino problem.  This problem is
essentially independent of the  previously-discussed solar
neutrino problems, since it depends strongly 
upon the $p$-$p$ neutrinos that are not observed in
the other experiments and whose calculated flux is approximately
model-independent.

The missing $^7$Be neutrinos cannot be
explained away by any change in solar physics. The \hbox{$^{8}$B}
neutrinos that are observed in the Kamiokande experiment are produced
in competition with the missing $^7$Be neutrinos; 
the competition is between electron capture on $^7$Be versus
proton capture on $^7$Be.
Solar model
explanations that reduce the predicted ${\rm ^7Be}$ flux 
generically reduce much
more (too much) the predictions for the observed ${\rm ^8B}$ flux.

The flux of $^7$Be neutrinos, $\phi({\rm ^7Be})$, is independent of
measurement uncertainties in  the
cross section for the nuclear reaction
${\rm ^7Be}(p,\gamma)^8$B; the  cross
section for this proton-capture  reaction is  the most uncertain
quantity that enters  in an important way in the solar
model calculations.  The flux of $^7$Be neutrinos depends upon the
proton-capture
reaction only through the ratio
\begin{equation}
\phi({\rm ^7Be}) ~\propto~ {{R(e)} \over {R(e) + R(p)}} ,
\label{Beratio}
\end{equation}
where $R(e)$ is the rate of electron capture by $^7$Be nuclei and
$R(p)$ is the rate of proton capture by $^7$Be.  With standard
parameters, solar models yield $R(p) \approx 10^{-3} R(e)$.
Therefore, one would have to increase the value of
the ${\rm ^7Be}(p,\gamma)^8$B cross section
by more than 2 orders of magnitude over the current best-estimate
(which has an estimated uncertainty of \hbox{$\sim$  10\%}) in order to affect
significantly the calculated $^7$Be solar neutrino flux.
The required change in the nuclear physics cross section
 would also  increase the predicted neutrino event
rate by more than 100 in the Kamiokande experiment, making that
prediction completely inconsistent with what is observed.
(From time to time, papers have been published claiming to solve the
solar neutrino problem by artificially changing the rate of the $^7$Be
electron capture reaction.
Equation~(\ref{Beratio}) shows that the flux of $^7$Be neutrinos
is actually independent of the rate of the electron capture reaction
to an accuracy of  better than 1\%.)

I conclude that either: 1) 
at least three of the four operating solar neutrino
experiments (the two gallium experiments plus either chlorine or
Kamiokande) 
 have yielded misleading results, or 2) physics beyond the standard
electroweak model is required to change the neutrino energy spectrum (or
flavor content) after the neutrinos are produced in the center of the sun.

\section{``The Last Hope'': No Solar Model}
\label{lasthope}

The clearest way to see that the results of the four solar neutrino
experiments are inconsistent with the predictions of the minimal
electroweak model is not to use standard solar models at all in the
comparison with observations. 
This is what Berezinsky, Fiorentini, and
Lissia~\cite{Berezinsky96} have termed ``The Last Hope'' for a solution
of the solar neutrino problems without introducing new physics.

Let me now explain how model independent tests are made.

Let $\phi_i(E)$ be the normalized shape of the neutrino energy
spectrum from one of the $i$ neutrino sources in the sun (e.g., 
$^8$B or $p-p$ neutrinos). I have shown~\cite{Bahcall91} that the shape
of the neutrino energy spectra that result from radioactive decays, 
 $^8$B, $^{13}$N, $^{15}$O, and $^{17}$F, are the same 
to $1$ part in $10^5$ as the laboratory shapes.  The $p-p$ neutrino
energy spectrum, which is produced by fusion  has a slight dependence
on the solar temperature, which affects the shape by about $1$\%. 
The energies of the neutrino lines from $^7$Be and $pep$ 
electron capture reactions are
also only slightly shifted, by about  $1$\% or less, 
because of the thermal energies of particles in the solar core.  

Thus one can test the hypothesis that an arbitrary linear combination
of the normalized neutrino spectra,

\begin{equation}
\Phi(E) ~=~\sum_{i} \alpha_i \phi_i(E),
\label{arbitrary}
\end{equation}
can fit the results of the neutrino experiments.
One can add a constraint to Eq.~(\ref{arbitrary}) that embodies the
fact that the sun shines by nuclear fusion reactions that also produce
the neutrinos.  
The explicit form of this luminosity constraint is 
\begin{equation}
\frac{L_\odot}{4\pi r^2} = \sum_j \beta_j \phi_j~,
\label{explicit} 
\end{equation}
where the eight coefficients, $\beta_j$, are given in
Table~VI of the paper by Bahcall and Krastev~\cite{BK96}.

The first demonstration that the four pioneering experiments are
by themselves inconsistent with the assumption that nothing happens to
solar neutrinos after they are created in the core of the sun was by 
Hata, Bludman,
and Langacker~\cite{Hata94}.  
They showed that the solar neutrino data available by late 1993 were
incompatible with any solution of equations (\ref{arbitrary}) and 
(\ref{explicit}) at the 97\% C.L. 

In the most recent and complete analysis in which the neutrino fluxes
are treated as free parameters,
Heeger and Robertson~\cite{HR96} showed that the data
presented at the Neutrino '96 Conference in Helsinki are  inconsistent
with equations (\ref{arbitrary}) and 
(\ref{explicit}) at the 99.5\% C.L.   Even if they omitted the
luminosity constraint, equation (\ref{explicit}), they found
inconsistency at the 94\% C.L. 

It seems to me that these demonstrations are so powerful and general
that there is very little point in discussing potential ``solutions''
to the solar neutrino problem based upon hypothesized non-standard
scenarios for solar models. 

\section{Comparison with Helioseismological Measurements}
\label{helioseismology} 

Helioseismology has recently sharpened
the disagreement between observations and the predictions of 
solar models 
with standard (non-oscillating) neutrinos.
This development has occurred in two ways.  

Helioseismology has confirmed the correctness of including diffusion
in the solar models and the effect of diffusion leads to somewhat
higher predicted events in the chlorine and  Kamiokande solar neutrino
experiments~\cite{BP95}. Even more importantly, helioseismology has
demonstrated that the sound velocities predicted by 
standard solar models agree with extraordinary precision with the
sound velocities of the sun inferred from helioseismological
measurements~\cite{BPBJCD}.  Because of the precision of this
agreement, I am convinced that standard solar models cannot be in
error by enough to make a major difference in the solar neutrino
problems.

The physical basis for the helioseismological measurements was
described beautifully by J{\o}ergen Christensen-Dalsgaard in his talk
yesterday afternoon.  I recommend  the text of his
discussion that you will also find in these proceedings.
You will see in J{\o}ergen's article references to other papers about 
helioseismology that you can use to  become better 
acquainted with the
subject. 

I will report here on some comparisons that Marc Pinsonneault,
Sarbani Basu, J{\o}ergen, and I have done recently 
which demonstrate the precise
agreement between the sound velocities in standard solar models and
the sound velocities inferred from helioseismological measurements.
These results are based upon an article that has since appeared 
 in {\it
Physical Review Letters}~\cite{BPBJCD}.

Since the deep solar interior behaves essentially
as a fully ionized perfect gas, $c^2 \propto T/\mu$
where $T$ is temperature and $\mu$ is mean molecular weight.
The sound velocities in the sun are determined from helioseismology to
a very high accuracy, better than 
 $0.2$\% rms throughout nearly all the sun.
Thus even tiny fractional  errors in  the model  
values of $T$ or $\mu$
would produce measurable discrepancies  in the precisely determined
helioseismological sound speed
\begin{equation}
{\delta c \over c}  \simeq 
{1 \over 2} \left({\delta T \over T}  - {\delta \mu \over \mu }\right)
\; .
\label{deltac}
\end{equation}
The remarkable numerical 
agreement 
between standard predictions and helioseismological 
observations, which I will discuss in the following remarks, 
 rules out   solar models with 
temperature or mean molecular weight 
profiles that differ significantly from standard profiles.
The helioseismological data essentially rule out solar models in which
deep mixing has occurred (cf. {\it PRL} paper\cite{els90}) 
and argue against unmixed models in which the
subtle effect of particle diffusion--selective sinking
of heavier species in the sun's gravitational field--is not included.

\begin{figure}[t]
\centerline{\psfig{figure=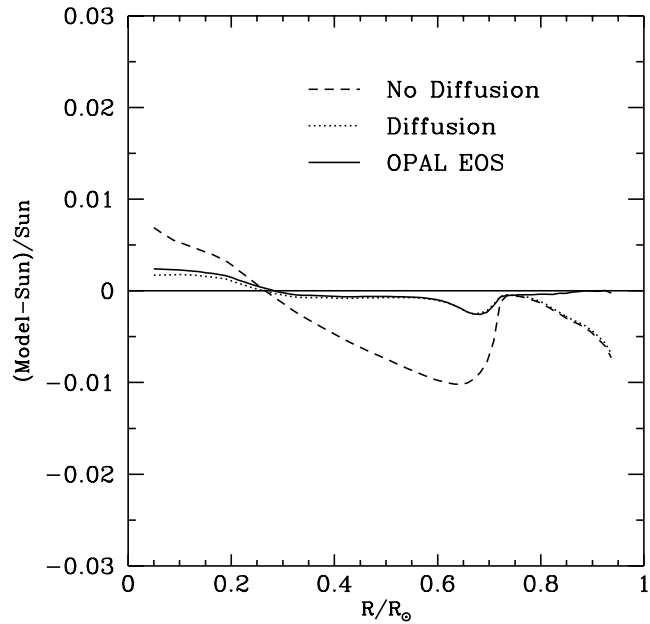,width=3.5in}}
\caption[]{Comparison of sound speeds predicted by different
standard solar models with the
sound speeds measured by  helioseismology.  
There are no free parameters in the models; the microphysics is
successively improved by first including diffusion and then by using
a more comprehensive equation of state.
The figure shows the
fractional difference, $\delta c/c$, 
between the predicted model sound speed and the 
measured~\cite{Basu96a,Basu96b}
solar values as a function of radial position in the sun
($R_\odot$ is the solar radius).
The dashed line refers to a model~\cite{BP95} in which
diffusion is neglected  and the dotted line was computed
from a model~\cite{BP95} in which helium and heavy element diffusion are 
included.
The dark line represents a model which includes 
recent improvements in the 
OPAL equation of state and opacities~\cite{Opacity,eos}.\label{fig:one}}
\end{figure}

Figure~\ref{fig:one} compares the  sound speeds computed
from three different solar models with the values
inferred~\cite{Basu96a,Basu96b} from the helioseismological measurements.
The 1995 standard model of Bahcall and Pinsonneault (BP)~\cite{BP95},
which includes  helium and heavy element diffusion, is represented by
the dotted line; the corresponding BP model 
without diffusion is represented by the dashed
line. The dark line represents the  best solar model which includes
recent improvements~\cite{Opacity,eos} in the OPAL equation of state and
opacities, as well as helium and heavy element diffusion.
For the OPAL EOS model,
the rms discrepancy  
between  predicted and measured sound speeds
is  $0.1$\% (which may be due partly to systematic uncertainties in the
data analysis).

In the outer parts of the sun, in the convective region between 
$0.7 R_\odot$ to $0.95 R_\odot$ (where the measurements end), the
No Diffusion and the 1995 Diffusion model have 
discrepancies as large as $0.5$\% (see Figure~\ref{fig:one}).  
The model with the Livermore
equation of state~\cite{eos}, OPAL EOS, fits the observations remarkably
well in this region.  We conclude, in agreement with the work of other
authors~\cite{Guenther96}, that the OPAL (Livermore
National Laboratory) equation of state provides 
a significant improvement in the description of
the outer regions of the sun.

The agreement between standard models and solar observations
is independent of the finer details of the solar model.  
The standard model of Christensen-Dalsgaard {\it et al.}~\cite{science96},
which is derived from an independent computer code with  different
descriptions of the microphysics, 
predicts solar sound speeds that agree everywhere with the measured
speeds to better than $0.2$\%.

Figure~\ref{fig:one} shows that the discrepancies with the No
Diffusion model are as large as 
$1$\%.  
The mean squared discrepancy for the No Diffusion model is 22 times
larger than for the best model with diffusion, OPAL EOS.  If one
supposed optimistically that the No Diffusion model were correct, one
would have to explain why the diffusion model fits the data so much
better. 
On the basis of Figure~\ref{fig:one}, we conclude that otherwise
standard solar
models that do not include diffusion, such  as the model  of 
Turck-Chi\`eze 
and Lopez~\cite{TC93}, 
are inconsistent with helioseismological observations.
This conclusion is consistent with earlier inferences based upon 
comparisons with  less complete
helioseismological data~\cite{Basu96a,jcd93,els90}, including the fact that
the present-day surface helium abundance in a standard solar 
 model agrees with
observations only if diffusion is included~\cite{BP95}.

Equation~\ref{deltac} and Figure~\ref{fig:one} imply that any changes
$\delta T/T$ from the standard model values of  temperature 
must be almost exactly canceled  by 
changes $\delta \mu/\mu$ in mean molecular weight.
In the standard model, $T$ and $\mu$ vary, respectively, by a factor
of $53$ and $43$\% over the entire range for which $c$ has been
measured and by $1.9$ and $39$\% over the energy producing region.
It would be a remarkable coincidence if nature chose $T$ and $\mu$
profiles that individually differ markedly from the standard model but
have the same ratio $T/\mu$.
Thus we expect 
that the fractional differences between the solar and the model
temperature, $\delta T/T$, or mean molecular weights, $\delta \mu/\mu$,
are of similar magnitude to $\delta c^2/c^2$, i.e. (using the larger 
rms error, $0.002$, for the solar interior),

\begin{equation}
\vert \delta T/T \vert, ~\vert \delta \mu/\mu \vert ~ \ltorder ~ 0.004 .
\label{inequality}
\end{equation}

How significant for solar neutrino studies 
is the agreement between observation and prediction
that is shown in Figure~\ref{fig:one}? 
The calculated neutrino fluxes 
depend upon the central
temperature of the solar model 
approximately as a power of the temperature, 
${\rm Flux} \propto T^n$, where  for standard models 
the exponent $n$ varies 
from $n \sim -1.1$ for
the  $p-p$ neutrinos to  $n \sim +24$ for the $^8$B 
neutrinos~\cite{BU96}.  
Similar temperature scalings are found for non-standard solar 
models~\cite{Castellani94}. Thus,
 maximum temperature differences of
$\sim 0.2\%$ would produce changes in the different neutrino
fluxes of several percent or less, much less than 
required~\cite{newphysics} to
ameliorate the solar neutrino problems.

\begin{figure}[thb]
\centerline{\psfig{figure=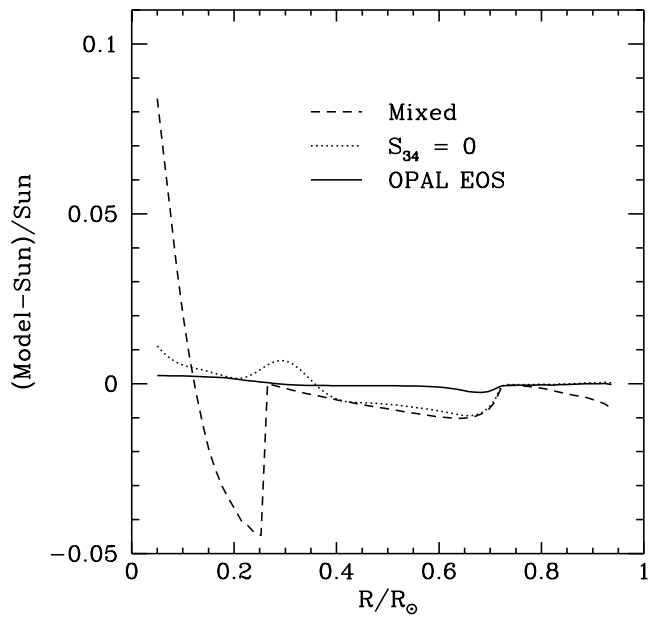,width=3.5in}}
\vskip -0.2in
\hglue-.2in\caption[]{Non-standard solar models compared with 
helioseismology.
This figure is similar to Figure \ref{fig:one} except that the
vertical scale is expanded. The dashed curve represents the sound
speeds computed for the mixed solar model of Cumming and
Haxton~\cite{Haxton96}  with
$^3$He mixing.  The dotted line represents the sound speed for a
solar
model computed with the rate of the 
$^3{\rm He}(\alpha,\gamma)^7$Be reaction 
set equal to zero.  
For comparison, we also include the results for the
new standard model labeled OPAL EOS in Figure~\ref{fig:one}.
\label{fig:two}}
\end{figure}

Figure~\ref{fig:two} shows that 
the ``mixed'' model of Cummings and Haxton (CH)~\cite{Haxton96}
(illustrated in  their Figure~1) is  grossly
inconsistent with the observed helioseismological measurements.
The vertical scale of Figure~\ref{fig:two} had to be 
expanded by a factor of $2.5$ relative to Figure~\ref{fig:one} in order 
to display the large discrepancies with observations for the mixed model.
The discrepancies for the CH mixed model
(dashed line in Figure~\ref{fig:two})
range from $+8$\% to $-5$\%.
Since $\mu$ in a standard solar model 
decreases monotonically outward from
the solar interior, the mixed model--with a constant value of $\mu$--
predicts too large values for the sound speed in the inner mixed
region and too small values in the outer mixed region.
The asymmetric form of the discrepancies for the CH model is due
to the competition between the assumed constant rescaling of the
temperature in the BP No Diffusion model and the assumed 
 mixing of the
solar core (constant value of $\mu$).
We also show in Figure~\ref{fig:two} the relatively
tiny discrepancies found for the new standard
model, OPAL EOS.   

More generally, 
helioseismology rules out all solar models with large amounts of
interior mixing, unless finely-tuned compensating changes in the
temperature are made.  The mean molecular weight
in the standard solar model with diffusion varies monotonically 
from $0.86$ in the
deep interior to $0.62$ at the outer region of nuclear fusion 
($R = 
0.25 R\odot$) to $0.60$ near the solar surface.
Any mixing
model will cause $\mu$ to be constant and equal to
the average value in the mixed region.
At the very least, 
the region in which nuclear fusion occurs must  be mixed 
in order to 
affect significantly the calculated neutrino 
fluxes~\cite{Bahcall89,EC68,BBU68,SS68,Schatzman69}.
Unless almost precisely canceling temperature changes are assumed,
solar models in which the nuclear burning region is mixed ($R \ltorder
0.25 R_{\odot}$)
will give maximum  differences, $\delta c$, between
the mixed and the 
standard model predictions, and hence between the mixed model
predictions and the observations, of order
\begin{equation}
{\delta c \over c} ~=~
{1 \over 2} \left({ {\mu - < \mu >} \over {\mu} }\right) ~\sim~ 7\%
~{\rm to}~ 10\%,
\label{maximum}
\end{equation}
which is inconsistent with Figure~\ref{fig:one}.

Are the helioseismological measurements sensitive to the rates of the
nuclear fusion reactions?  In order to answer this question in its
most extreme form, we have
computed a  model in which the cross section factor, $S_{34}$,
 for the 
$^3{\rm He}(\alpha, \gamma)^7{\rm Be}$ 
reaction is artificially set equal to zero. 
The neutrino fluxes computed from this unrealistic model 
have been  used~\cite{Bahcall89}
  to set a lower limit on the allowed 
rate of solar neutrinos in the gallium experiments if
the solar luminosity is currently powered by nuclear fusion reactions. 
Figure~\ref{fig:two} shows that 
although the maximum 
discrepancies ($\sim 1$\%) for  the $S_{34} = 0$ model
are much smaller than for mixed models,
they are still large compared to the differences between  the standard
model and 
helioseismological measurements.
The mean squared discrepancy for the $S_{34} = 0$ model is 19 times
larger than for the standard OPAL EOS model.
We conclude that the $S_{34} = 0$ model is not compatible with
helioseismological observations.

To me, these results suggest strongly that the assumption on
which they are based---nothing happens to the neutrinos after they are
created in the interior of the sun---is incorrect.  

\section{$^3$He Mixing versus MSW Mixing}
\label{versus}

It is instructive to compare  the success of the hypothesis of
$^3$He solar mixing\cite{Haxton96}  with the success of 
 the hypothesis of MSW neutrino mixing. I do
so below.

\bigskip
$\bullet${\bf Consistency With Solar Neutrino Experiments}.  The
$^3$He mixing hypothesis is inconsistent at the 99.5\% C.L. with 
solar neutrino experiments (a special case of the general result of
Heeger and Robinson\cite{HR96}); MSW mixing is consistent with 
the solar neutrino experiments (best value of $\chi^2$ is less than
one per degree of freedom).

\medskip
$\bullet${\bf Consistency With Helioseismology}.  The $^3$He mixing
hypothesis is inconsistent with helioseismology; mixing of the solar
core necessarily implies $\sim 7$\% discrepancies with the measured
solar sound velocities. The standard solar model with no free
parameters predicts sound velocities that agree with the measured
velocities to a rms accuracy of 0.2\% in the solar core.

\medskip
$\bullet${\bf Free Parameters}. The $^3$He mixing has 3 free
parameters; the MSW mixing has 2 free parameters.

\section{Discussion}
\label{closing}

The combined predictions of the standard solar model and the standard
electroweak theory disagree with the results of the four pioneering
solar neutrino experiments.  The disagreement persists even if the 
neutrino fluxes are treated as free parameters, without reference to 
any solar model.

The  solar model calculations are in excellent agreement with 
helioseismological measurements of the sound velocity, providing
further support for the inference that something happens to the solar
neutrinos after they are created in the center of the sun.

Looking back on what was envisioned in 1964, I am astonished and
pleased with what has been accomplished. In 1964, it was not clear 
that solar neutrinos could be detected.  Now, they have been observed
in different experiments and the theory of stellar energy generation by
nuclear fusion has been directly confirmed. Moreover, particle
theorists have shown that solar neutrinos can be used to study
neutrino properties, a possibility that we did not even consider in
1964.  In fact, much of the interest in the subject stems from the
fact that 
the four pioneering experiments suggest that new neutrino physics may 
be revealed by solar neutrino measurements.
Finally, helioseismology has confirmed to high precision predictions
of the standard solar model, a possibility that  also was not imagined
in 1964.

\section*{Acknowledgments}
This research is supported in part by NSF
grant number PHY95-13835.

\end{document}